# Thickness-Dependent Morphologies of Gold on *N*-Layer Graphenes

Haiqing Zhou,[†,§] Caiyu Qiu,[†,§] Zheng Liu,[†,§] Huaichao Yang,[†,§] Lijun Hu,[†,§] Ji Liu,[†,§] Haifang Yang,[‡] Changzhi Gu,[‡] and Lianfeng Sun*,[†]

*National Centre for Nanoscience and Technology, Beijing, 100190, China, Beijing National Laboratory for Condensed Matter Physics, Institute of Physics, Chinese Academy of Sciences, Beijing, 100190, China, and Graduate School of Chinese Academy of Sciences, Beijing, 100049, China*

Received November 4, 2009; E-mail: slf@nanoctr.cn

Since their discovery in 2004,[1] *n*-layer graphenes have attracted intense interest due to their unique structure; interesting, novel properties;[2,3] and significant dependence of properties on the layer number "*n*".[4,5] For example, in monolayer and bilayer graphene, the carriers correspond to massless and massive Dirac fermions and the quantum hall effect occurs at different filling factors, respectively.[4,5] In this work we report that gold deposited by thermal evaporation on graphenes interact differently with these substrates depending on the layer number of graphenes. This results in different morphologies of gold film on graphenes with different layer numbers. The differences observed with SEM (scanning electron microscopy) can be used to identify and distinguish graphenes especially if they are mixed with different layer numbers. Compared to the usual two methods reported previously, this technique with SEM is more convenient, has a higher throughput than that based on AFM (atomic force microscopy),[1] and has a higher spatial resolution than that based on Raman spectroscopy.[6−9]

The *n*-layer graphenes investigated in this work were prepared with the micromechanical cleavage[1] of natural graphite (Alfa Aesar) by using Scotch transparent tape 600 (3M).[10] After repeated peeling, the graphenes were transferred to 300 nm SiO$_2$/Si substrates by adhering and taking off the tape. The *n*-layer graphenes left on the wafer were first selected based on their color contrast in an optical microscope (Leica DM4000), and their coordinates were found with regard to the marks in wafer. The Raman spectra of these *n*-layer graphenes were used to identify the layer number (Figure 1).[6−9] The micro-Raman spectroscopy (Renishaw inVia Raman Spectroscope) experiments were performed under ambient conditions with 514.5 nm (2.41 eV) excitation from an argon ion laser. The laser power on the sample was ∼1.0 mW to avoid laser induced heating. A 100× objective lens with a numerical aperture of 0.90 was used, which resulted in an ∼1 μm laser spot size.

To study the surface properties of *n*-layer graphenes,[3,11] gold film was evaporated onto the wafer in a vacuum thermal evaporator at a deposition rate of 1.0 Å/s under a vacuum of ∼10$^{-4}$ Pa). The morphologies of gold film as well as the size and density of nanoparticles after heat treatment were studied in detail by SEM (Figure 2).

As shown in Figure 3, the morphologies of gold film on *n*-layer graphenes are closely related to the layer number "*n*". The grain size of gold and aperture among the grains are larger for the gold film on bilayer graphene compared to that on monolayer graphene, which results in a rougher surface of gold on bilayer graphene (Figure 3a). Thus, a boundary can be found between the gold film on monolayer and bilayer graphene. The differences in morphologies, grain size, and apertures among grains can be more clearly seen from Figure 3b, where the left and right parts correspond to gold films on trilayer and monolayer graphene, respectively. Here the straight boundary is clearer. In Figure 3c, the morphologies of gold film are shown on the four layer graphene, substrate (SiO$_2$), and bilayer graphene, respectively. In Figure 3d, a narrow ribbon of bilayer graphene in the left edge of monolayer graphene can be clearly seen and identified, which would be taken to be monolayer graphene by an optical microscope and Raman identification (see Supporting Information). The high spatial resolution of this technique can be further demonstrated by the circles in Figure 3, which are defects or contaminants on *n*-layer graphenes.

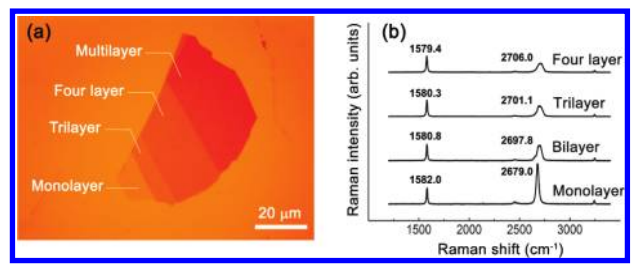

**Figure 1.** (a) A typical photograph (in normal white light) of an *n*-layer graphenes. The layer number is estimated according to the color contrast. (b) Raman spectra of pristine *n*-layer graphenes, which can be used to confirm the layer number. The left and right peaks correspond to G and 2D bands, respectively.

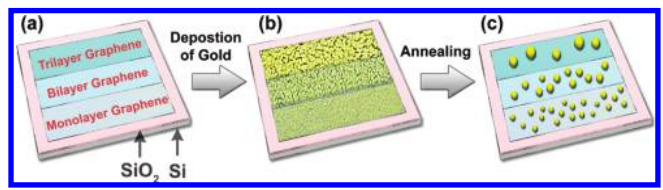

**Figure 2.** Schematic diagram showing the technique to study the surface properties of *n*-layer graphenes using gold film or nanoparticles. (a) Graphenes with different layer numbers are found. (b) A thin film of gold or other metals is evaporated onto the wafer, and the morphologies on *n*-layer graphenes were studied by SEM (c) The wafer is treated at high temperature, and the size, density, etc. of nanoparticles on *n*-layer graphenes can be obtained.

What is the mechanism of these thickness-dependent morphologies of gold on *n*-layer graphenes? Generally speaking, the final morphologies of gold are closely related to thermodynamic (e.g., energetics and stability) and kinetic (e.g., surface diffusion) factors.

---

[†] National Centre for Nanoscience and Technology.
[§] Graduate School of Chinese Academy of Sciences.
[‡] Beijing National Laboratory for Condensed Matter Physics, Chinese Academy of Sciences.

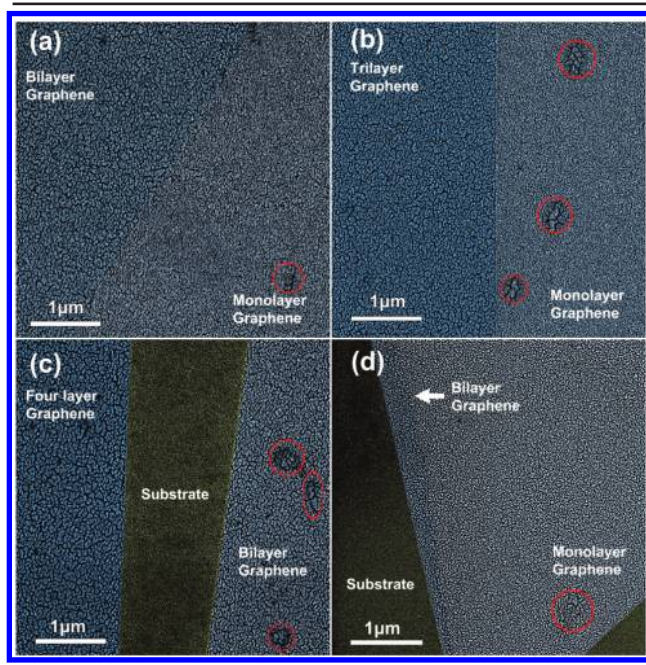

*Figure 3.* SEM images of the morphologies of gold film on *n*-layer graphenes (false-color). Differences in morphologies of gold film and clear boundary can be found. The circles indicate defects or contaminants on graphenes. If the layer number is larger than 4, the difference is almost indistinguishable. Vacuum: $10^{-4}$ Pa. Deposition rate: 1.0 Å/s. Film thickness: 5.0 nm in (a, b, c) and 2.0 nm in (d). (a) Gold film on bilayer (left) and monolayer graphene (right). (b) Gold film on trilayer (left) and monolayer graphene (right). (c) Gold film on four layer graphene (left), substrate ($SiO_2$, middle), and bilayer graphene (right). (d) Gold film on substrate, a narrow bilayer, and monolayer graphene. It is difficult to distinguish the bilayer graphene from the monolayer graphene using an optical microscope or Raman spectra, indicating the advantage of identifying the layer number with SEM.

One possible reason may be that the surface free energy of *n*-layer graphenes is dependent on the layer number. When the gold atoms are evaporated onto the surface of *n*-layer graphenes, they interact differently and result in different morphologies.

Another possible reason can be attributed to a kinetic factor: surface diffusion. When gold atoms are thermally deposited onto the silicon wafer, the arriving atoms can make random walks on the surface ($SiO_2$ or graphenes; see Supporting Information). An arriving atom may form a new island with other atoms (nucleation) or walk into an existing one (growth).[12] The diffusion coefficient of the atoms determines how large an area an atom interrogates in unit time. Therefore, the diffusion coefficient determines the outcome of the competition between nucleation and growth and the number density of islands after deposition.[13] A large diffusion coefficient means a high probability for an arriving atom to find an existing island, yielding a low number of islands after deposition. Therefore, different surface diffusion coefficients can result in thickness-dependent morphologies, grain sizes, and apertures of gold film on *n*-layer graphenes.

The different morphologies of gold on *n*-layer graphenes can be demonstrated more clearly after heat treating. The silicon wafer, covered with graphenes and gold films, is placed in a molybdenum boat using direct current heating, and the temperature is monitored with an infrared thermometer. Figure 4 shows a typical SEM image of the morphologies of gold nanoparticles on *n*-layer graphenes after annealing at 1260 °C for 30 s. It is evident from these images that the size and density of nanoparticles are quite different on graphenes with different layer numbers. The average size of gold nanoparticles on monolayer graphene is the smallest, while the density is the highest. For the gold nanoparticles on bilayer graphene, their average size is larger than that on monolayer graphene and smaller than that on trilayer graphene (Figure 4a). Meanwhile, the density of gold nanoparticles on bilayer graphene is smaller than that on monolayer graphene and larger than that on trilayer graphene.

To obtain a statistical average grain size and density of gold nanoparticles on *n*-layer graphenes, we have measured the sizes and counted the number of nanoparticles except the edge area from Figure 4a and several other images, and the results are shown in Figure 4b. The average size is $32.6 \pm 13.3$, $50.1 \pm 22.2$, $65.7 \pm 27.0$, $80.8 \pm 33.3$ nm for gold nanoparticles on monolayer, bilayer, trilayer and four layer graphene, respectively. The average density of nanoparticles is $(9.23 \pm 0.50) \times 10^{13}$ /m$^2$ ($N_1$), $(2.60 \pm 0.24) \times 10^{13}$ /m$^2$ ($N_2$), $(1.25 \pm 0.15) \times 10^{13}$/m$^2$ ($N_3$), $(0.68 \pm 0.10) \times 10^{13}$/m$^2$ ($N_4$) on monolayer, bilayer, trilayer, and four layer graphene, respectively.

Since the heat treating temperature is well over the melting temperature of gold, it is reasonable to assume that the gold nanoparticles are spherical. Thus the total mass of gold that still remains on the *n*-layer graphene can be calculated after annealing (see Supporting Information). There are still 49.4%, 55.2%, 56.0%, and 58.6% left on monolayer, bilayer, trilayer and four layer graphene, respectively. Meanwhile, just several gold nanoparticles can be found occasionally on the $SiO_2$ substrate (Figure 4a). The absence of gold nanoparticle on the substrate and loss of mass on *n*-layer graphene can be attributed to the evaporation of gold atoms into the vacuum because of the high temperature treatment. These results indicate that the adhesion of gold with *n*-layer graphene is stronger than that with $SiO_2$; and the adhesion of gold with *n*-layer graphenes varies with the thickness of graphenes.

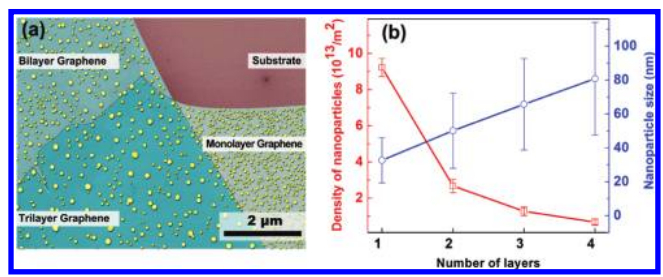

*Figure 4.* Morphologies, size, and density of gold nanoparticles on *n*-layer graphenes after annealing at 1260 °C in vacuum for 30 s (false-color). Note that no gold nanoparticles are found in the substrate. (a) Gold nanoparticles on monolayer, bilayer, and trilayer graphene, respectively. (b) Statistics of the size and density of gold nanoparticles on *n*-layer graphenes.

The nanoparticle density of gold on *n*-layer graphene is closely related to the surface diffusion coefficient (*D*) according to classical mean-field nucleation theory.[12,13] The scaling relationship between the nanoparticle density (*N*) and the surface diffusion coefficient (*D*) can be expressed as: $N \propto (1/D)^{1/3}$ for isotropic surface diffusion. According to the Arrhenius equation of diffusion, the surface diffusion coefficient and the diffusion barriers ($E_n$) follow the scaling relation of $D \propto \exp(-E_n/KT)$, where $K$ is the Boltzmann constant, $T$ the temperature (see Supporting Information). Thus, combining these two equations, we can obtain: $N \propto \exp(E_n/3KT)$. Therefore, the diffusion barrier $E_n$ of gold is dependent on the layer number of the *n*-layer graphene. Although it is difficult to obtain the absolute value of barriers, the barrier difference between *n*-layer graphene can be estimated by the density ratios. Therefore, the barrier difference between monolayer and bilayer graphene can be calculated with $\Delta E = E_1 - E_2 = 3KT \ln(N_1/N_2) = 504 \pm 44$ meV. The barrier difference between other layer numbers can be obtained in a similar method, and they are 291 ± 31 meV ($E_2 - E_3$) and 242 ± 22 meV ($E_3 - E_4$), respectively.

What makes the surface diffusion coefficient and barrier of gold on *n*-layer graphene different? The quantum size effect (QSE)[14,15] can modulate the band gap of semiconducting nanocrystals, and hence they exhibit size-dependent visible color.[16] In ultrathin metal film, the electrons are confined in the vertical direction, and QSE is seen by the appearance of discrete energy levels of electrons.[17,18] For monolayer graphenes, the pi-electrons are confined and reside above and below the graphene layer. The stacking of graphene layers does not increase the confinement of electrons. Recently it is discovered that the tunable band gap can be opened in bilayer graphenes.[19−22] This suggests that the different surface diffusion coefficient and barrier of gold on *n*-layer graphenes may originate from QSE in a similar mechanism to that of semiconducting nanocrystals, which is closely related to van der Waals coupling between graphene layers.

We have tried to apply this methodology to graphenes obtained by reduction of graphene oxide.[23,24] Although the morphologies of gold film on graphenes are different from that on SiO$_2$, the morphologies of gold are not uniform and show little dependence on the thickness of graphenes due to the higher density of defects or chemical contaminants (see Supporting Information).

It is worth noting that the gold film can be completely evaporated off the *n*-layer graphenes if annealing for several minutes at temperature above the melting point of gold (see Supporting Information). Micro-Raman spectra indicate that the nature of graphenes is not altered after the high temperature treatment (Figure 5). Meanwhile, similar results have been obtained if gold is replaced with other metals, indicating the universal phenomena reported here (see Supporting Information).

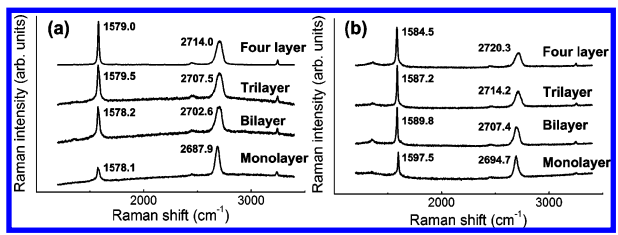

*Figure 5.* The micro-Raman spectra of the gold-covered and heat-treated *n*-layer graphenes. The left and right peaks correspond to G and 2D bands, respectively. (a) Raman spectra of *n*-layer graphenes covered with gold film of 5.0 nm. (b) Raman spectra of *n*-layer graphenes after annealing.

In summary, gold thermally deposited onto *n*-layer graphenes interacts differently with these substrates depending on the layer number, indicating the different surface properties of graphenes. This results in thickness-dependent morphologies of gold on *n*-layer graphenes, which can be used to identify and distinguish graphenes with high throughput and spatial resolution. It plays an important role in checking if *n*-layer graphenes are mixed with different layer numbers of graphene with a smaller size, which cannot be found by Raman spectra and are deleterious to the electrical studies of *n*-layer graphenes.[25−27]

**Acknowledgment.** This work is supported by "973" Program of Ministry of Science and Technology (2006CB932402) and National Science Foundation of China (Grant No. 10774032, 90921001, 50952009). We thank Prof. B. H. Han and Dr. S. Z. Zu for graphene oxide samples.

**Supporting Information Available:** The results of the corresponding optical images of Figure 3, a model of gold atom on graphenes, the calculation methods of the total volume of gold left on graphenes, a model to show the different diffusion barriers of gold on graphenes, gold film on graphenes obtained by reduction of graphene oxide, SEM and EDX showing that the gold film can be evaporated off graphenes, primary results with Cu, Ag nanoparticles on graphenes are given. This material is available free of charge via the Internet at http://pubs.acs.org.

**References**

(1) Novoselov, K. S.; Geim, A. K.; Morozov, S. V.; Jiang, D.; Zhang, Y.; Dubonos, S. V.; Grigorieva, I. V.; Firsov, A. A. *Science* **2004**, *306*, 666.
(2) Zhang, Y. B.; Tan, Y. W.; Stormer, H. L.; Kim, P. *Nature* **2005**, *438*, 201.
(3) Geim, A. K. *Science* **2009**, *324*, 1530.
(4) Geim, A. K.; Novoselov, K. S. *Nat. Mater.* **2007**, *6*, 183.
(5) Novoselov, K. S.; Geim, A. K.; Morozov, S. V.; Jiang, D.; Katsnelson, M. I.; Grigorieva, I. V.; Dubonos, S. V.; Firsov, A. A. *Nature* **2005**, *438*, 197.
(6) Gupta, A.; Chen, G.; Joshi, P.; Tadigadapa, S.; Eklund, P. C. *Nano Lett.* **2006**, *6*, 2667.
(7) Ferrari, A. C.; Meyer, J. C.; Scardaci, V.; Casiraghi, C.; Lazzeri, M.; Mauri, F.; Piscanec, S.; Jiang, D.; Novoselov, K. S.; Roth, S.; Geim, A. K. *Phys. Rev. Lett.* **2006**, *97*, 187401.
(8) Ni, Z. H.; Wang, H. M.; Ma, Y.; Kasim, J.; Wu, Y. H.; Shen, Z. X. *ACS Nano* **2008**, *2*, 1033.
(9) Calizo, I.; Balandin, A. A.; Bao, W.; Miao, F.; Lau, C. N. *Nano Lett.* **2007**, *9*, 2645.
(10) Li, Q.; Li, Z. J.; Chen, M. R.; Fang, Y. *Nano Lett.* **2009**, *9*, 2129.
(11) Wang, X. R.; Tabakman, S. M.; Dai, H. J. *J. Am. Chem. Soc.* **2008**, *130*, 8152.
(12) Mo, Y. W.; Kleiner, J.; Webb, M. B.; Lagally, M. G. *Phys. Rev. Lett.* **1991**, *66*, 1998.
(13) Ma, L. Y.; Tang, L.; Guan, Z. L.; He, K.; An, K.; Ma, X. C.; Jia, J. F.; Xue, Q. K. *Phys. Rev. Lett.* **2006**, *97*, 266102.
(14) Ekimov, A. I.; Efros, A. L.; Onushchenko, A. A. *Solid State Commun.* **1985**, *56*, 921.
(15) Valden, M.; Lai, X.; Goodman, D. W. *Science* **1998**, *281*, 1647.
(16) Colvin, V. L.; Schlamp, M. C.; Alivisatos, A. P. *Nature* **1994**, *370*, 354.
(17) Guo, Y.; Zhang, Y. F.; Bao, X. Y.; Han, T. Z.; Tang, Z.; Zhang, L. X.; Zhu, W. G.; Wang, E. G.; Niu, Q.; Qiu, Z. Q.; Jia, J. F.; Zhao, Z. X.; Xue, Q. K. *Science* **2004**, *306*, 1915.
(18) Ma, X. C.; Jiang, P.; Qi, Y.; Jia, J. F.; Yang, Y.; Duan, W. H.; Li, W. X.; Bao, X. H.; Zhang, S. B.; Xue, Q. K. *Proc. Natl Acad. Sci. U.S.A.* **2007**, *104*, 9204.
(19) Ohta, T.; Bostwick, A.; Seyller, T.; Horn, K.; Rotenberg, E. *Science* **2006**, *313*, 951.
(20) Castro, E. V.; Novoselov, K. S.; Morozov, S. V.; Peres, N. M. R.; Dos Santos, J. M. B. L.; Nilsson, J.; Guinea, F.; Geim, A. K.; Neto, A. H. C. *Phys. Rev. Lett.* **2007**, *99*, 216802.
(21) Zhou, S. Y.; Gweon, G. H.; Fedorov, A. V.; First, P. N.; De Heer, W. A.; Lee, D. H.; Guinea, F.; Neto, A. H. C.; Lanzara, A. *Nat. Mater.* **2007**, *6*, 770.
(22) Zhang, Y. B.; Tang, T. T.; Girit, C.; Hao, Z.; Martin, M. C.; Zettl, A.; Crommie, M. F.; Shen, Y. R.; Wang, F. *Nature* **2009**, *459*, 820.
(23) Goncalves, G.; Marques, P. A. A. P.; Granadeiro, C. M.; Nogueira, H. I. S.; Singh, M. K.; Gracio, J. *Chem. Mater.* **2009**, *21*, 4796.
(24) Zu, S. Z.; Han, B. H. *J. Phys. Chem. C* **2009**, *113*, 13651.
(25) Kosynkin, D. V.; Higginbotham, A. M.; Sinitskii, A.; Lomeda, J. R.; Dimiev, A.; Price, B. K.; Tour, J. M. *Nature* **2009**, *458*, 872.
(26) Jiao, L. Y.; Zhang, L.; Wang, X. R.; Diankov, G.; Dai, H. J. *Nature* **2009**, *458*, 877.
(27) Zhang, Z. X.; Sun, Z. Z.; Yao, J.; Kosynkin, D. V.; Tour, J. M. *J. Am. Chem. Soc.* **2009**, *131*, 13460.

JA909228N